# Picometer-scale atom position analysis in annular bright-field STEM imaging

Peng Gao[a,b,c], Akihito Kumamoto[a], Ryo Ishikawa[a], Nathan Lugg[a], Naoya Shibata[a], Yuichi Ikuhara[a,d]

[a]Institute of Engineering Innovation, School of Engineering, University of Tokyo, Tokyo 113-8656, Japan
[b]Electron Microscopy Laboratory, School of Physics, and Center for Nanochemistry, Peking University, Beijing100871, China
[c]Collaborative Innovation Center of Quantum Matter, Beijing100871, China
[d]Nanostructures Research Laboratory, Japan Fine Ceramic Center, Nagoya 456-8587, Japan

Email: p-gao@pku.edu.cn; ikuhara@sigma.t.u-tokyo.ac.jp

We study the effects of specimen mistilt on the picometer-scale measurement of local structure by combing experiment and simulation in annular bright-field scanning transmission electron microscopy (ABF-STEM). A relative distance measurement method is proposed to separate the tilt effects from the scan noise and scan distortion. We find that under a typical experimental condition a small specimen tilt (~6 mrad) in 25 nm thick $SrTiO_3$ along [001] causes 11.9 pm artificial displacement between O and Sr/TiO columns in ABF image, which is more than 3 times of scan noise and sample drift induced image distortion ~3.2 pm, suggesting the tilt effect could be dominant for the quantitative analysis of ABF images. The artifact depends the crystal mistilt angle, specimen thickness, defocus, convergence angle and uncorrected aberration. Our study provides useful insights into detecting and correcting tilt effects during both experiment operation and data analysis to extract the real structure information and avoid mis-interpretations of atomic structure as well as the properties such as oxygen octahedral distortion/shift.

## 1. Introduction

Local structure distortion at grain boundary, hetero-interface, dislocation and surface can significantly influence on a broad variety of physical properties in complex oxide materials. Precise measurement of atom positions at these defects enables us quantitatively analyze the local strain field, electric dipole, flexoelectric effects, and chemical valence and thus provide new insights into understanding as to how materials properties depend on the local atomic structures and how we can engineer defects to optimize the materials or devices. Traditional crystallographic structure analysis such as x-ray or neutron diffraction is inadequate for the local structure analysis because of the poor spatial resolution. The recent advances in aberration corrected (scanning) transmission electron microscopy (S/TEM), however, can allow us to directly measure the inter-atomic distances with picometer-precision, providing a unique tool to study the local structure distortion. For examples, Bals *et al.* measured the local atomic structure of $Bi_4W_{2/3}Mn_{1/3}O_8Cl$ by using exit wave reconstruction [1]. Jia *et al.* directly mapped the electric dipoles in ferroelectric thin films by using negative $C_s$ TEM observations [2]. The atomic displacements at ferroelectric domain walls in $BiFeO_3$ and $PbZr_{0.2}Ti_{0.8}O_3$ thin films are studied by measuring annular dark-field (ADF) STEM images from different groups [3, 4]. Yankovich *et al.* reported sub-picometer precision measurement of Si lattice in ADF-STEM imaging [5]. In TEM mode, the exit wave reconstruction or negative $C_s$ condition require systematic simulations for the interpretation of image contrast and also very thin specimens (typically less than 5 nm). For the atomic-resolution STEM imaging, the interpretation of image contrast is much straightforward and reliable [6-8], and it is also capable of simultaneous spectroscopy, being powerful to characterize the local structural and chemical properties. However, in the ADF-STEM images (Z-contrast, Z is atomic number), the functional light elements such as lithium or oxygen are usually invisible when the compounds contain relatively heavier elements. In contrast, the ABF imaging [8-12] are able to simultaneous visualize both heavy and light element atomic columns over a wide range of thickness (typically more than 50 nm), allowing us to determine the positions of full atomic species from a single image. Therefore, ABF-STEM imaging is one of best candidate for the quantitative measurement of local structure distortion, especially for the complex oxides analysis [13, 14].

Generally, the measurement precision of inter-atomic distance in STEM is limited by scan noise and image distortion originated from the specimen drift (we use 'scan distortion' hereinafter). Several methods, therefore, have been employed to correct scan distortion [5, 15-18]. In fact, the specimen tilt is another important factor that could influence on the contrast and deformation of the atom-shape, as reported in ADF-STEM [19-27] due to the reduction in strength of the electron beam channeling [28]. On the basis of dynamical scattering theory, Van Dyck *et al.* discussed the effects of different range of specimen tilt [19]. Maccagnano-Zacher *et al.* pointed out that a specimen tilt reduces the contrast in atomic-resolution ADF-STEM [20]. So *et al.* reported that a specimen misalignment can cause a shift of atomic columns [21]. In the ABF image, the contrast of light atomic columns are basically related to forward elastic scattering, whereas the contrast of heavy atomic columns are contributed from both thermal diffuse scattering (TDS) and elastic scattering components [8]. Unlike the TDS signal (incoherent), the elastic scattering (coherent) contrast is very sensitive to the specimen conditions. One example is that the intensity of ABF image oscillates along the thickness, which becomes significant for the thin specimen and/or light atomic columns [8]. Therefore, the tilt effect in ABF image contrast is likely more sensitive than in ADF images, which has been firstly pointed out by Findlay et al. based on the $SrTiO_3$ simulations [8]. Recently, Gao et al. reported that the crystal tilt might cause significant artifact in interpretation of the ferroelectric polarization phenomena in ABF images [13, 29]. Zhou et al. studied the deviation of atom positions between ABF and ADF and the effect of tilt on the bond angle measurements in $ZrO_2$ [30]. Liu et al. reported the effects of crystal tilt on the relative positions in $PbTiO_3$ in both ADF- and bright filed (BF) STEM images and found that tilt effect strongly depends on the tilt-angle and the specimen thickness [31]. Brown et al. proposed a new method to detect and correct the specimen tilt by simultaneously recording a central bright-field imaging (c-BF) and ABF images [32].

In this paper, we combine experiments and simulations to quantitatively study the small specimen tilt effects on the atom position analysis of ABF images with an emphasis on calculating and correcting the artifact, avoiding mis-interpretation of local atomic structure. Since the tilt is always mixed with scan noise and scan distortion in experimental images, we introduce a new method, relative-distance measurement, to separate the tilt effects from other factors. We choose perovskite $SrTiO_3$ ($ABO_3$) single crystal with perfect cubic structure as a demonstration to discuss the tilt effects on quantitative measurement of atomic column positions in ABF-STEM imaging. In a perovskite structure, the rigid oxygen octahedron exhibits various structural distortions driven by stain and electrostatic conditions, underpinning vast functionalities. Many efforts have been paid to precisely measure the local distortion in perovskite structures by various microscopy techniques [2-4, 26, 33-47] but not much in ABF-STEM imaging [13, 14, 29]. In our work, we use a probe corrected ARM300CF (JEOL Ltd.) microscope, operating at 300 kV, and we simultaneously record both ADF- and ABF-STEM images, where a convergence semi-angle is 24 mrad, detector collection semi-angles spanning from 65 to 200 mrad for ADF and 12 to 24 mrad for ABF imaging, respectively. We also have performed systematic multi-slice image simulations for quantitative comparison with experiment. Dynamical image simulations in STEM were carried out by using commercial software of HREM Research, Inc. The Debye-Waller factors for Sr, Ti and O are given in the previous literature [48].

We find that, in the tilted crystal, the atom position shift strongly depends on atom species, leading to significant artifacts in distance measurement between the light anion columns and heavy cation columns, i.e., artificial atomic displacements. Under typical convergence semi-angle of 24 mrad, even with the small tilt of 6 mrad (0.34°) with 25 nm thick $SrTiO_3$ viewing along the [001] direction, the artificial displacement is estimated to be 11.9 pm between cation and anions atoms, which must lead to significant mis-interpretation of local atomic structure. This artificial displacement is much larger than that induced by scan noise and distortion (a few picometers). This artificial displacements depend on the tilt angle, defocus, thickness of the specimen, convergence angle and uncorrected aberration. Under some certain experimental conditions, such artifacts can completely dominate the measurement error, even the tilt is as small as 0.5 mrad. Since such small specimen tilt is inevitable even under the deliberate experiment operation, it becomes critical to consider the tilt effect for the precise measurement of atom positions in ABF images. We present in details on mechanism of formation, estimation and correction of the artifact, providing useful insights into local structure measurements of ABF images with specimen tilt.

## 2. Method

The electron transparent specimen of $SrTiO_3$ were prepared by mechanical polishing the single crystal followed by Ar ion milling (Precision Ion Polishing System, Gatan). Figs. 1a and b are simultaneously and sequentially recorded ADF and ABF images along the [001] direction with sampling rate of 5 pm per pixel [49]. Generally, the coexist of specimen tilt and drift effects are difficult to separate. However, in the case of sequential imaging, the effect of specimen drift (scan distortion) can be minimized and the tilt effects stand out. Atomistic structure model is overlaid on the images and highlights the positions of Sr, TiO and O columns in Fig.1a and b. Fig. 1c is the contrast inverted ABF image (*I*-ABF) in which the atom positions and the contrast-deformation of atoms are easier to distinguish. Figs. 1d-f show the unit-repeated-averaged images to reduce the scan noise. Although the bulk of single crystal $SrTiO_3$ must have a perfect cubic structure without any oxygen octahedral distortion/shift, the O columns in Fig. 1f are not at the symmetric positions relative to cation columns, and instead shift to bottom-right direction due to the specimen tilt.

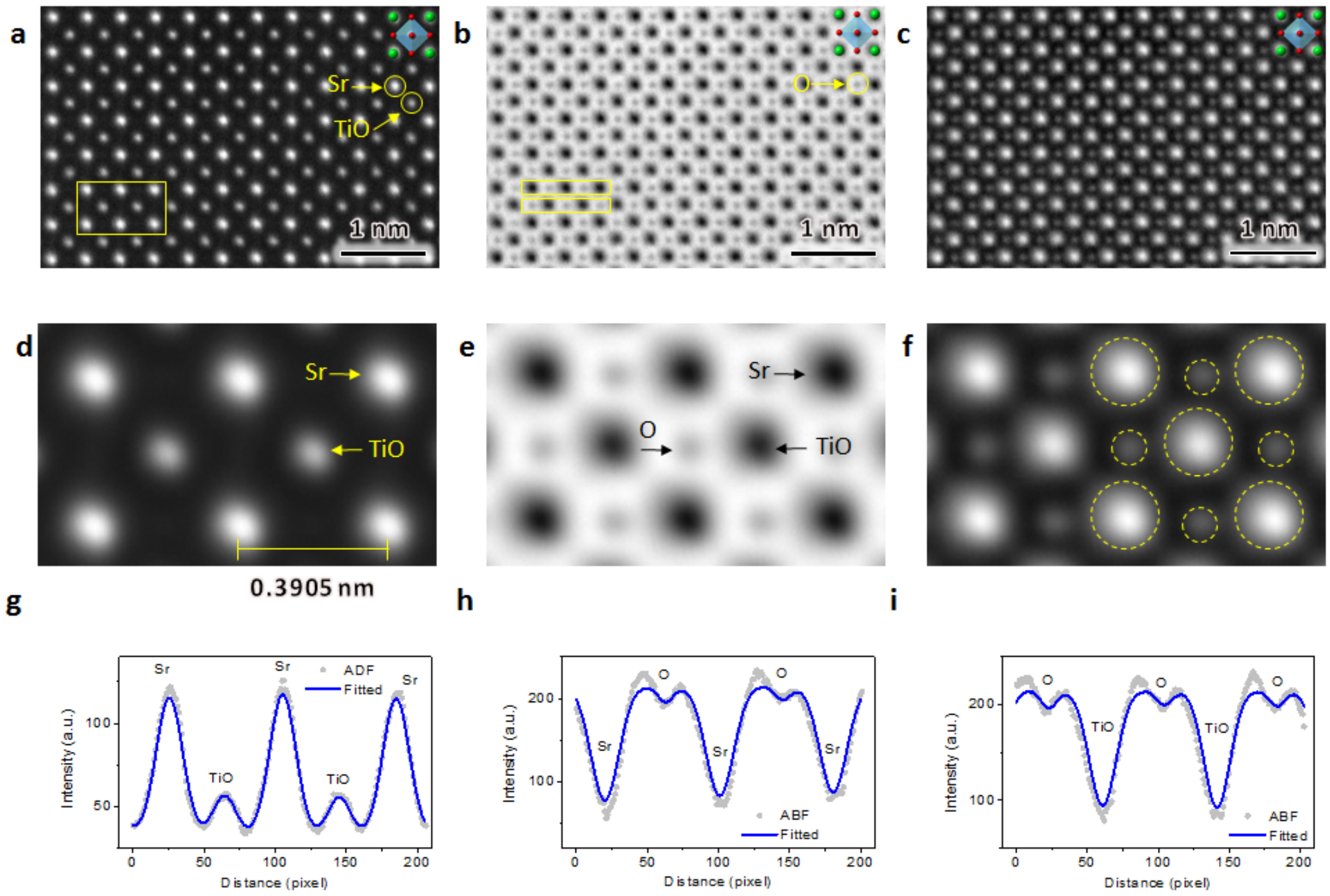

**Figure1.** (a) Raw ADF-STEM image from a single crystal $SrTiO_3$ (STO) with zone axis close to [0 0 1]. (b) Simultaneously recorded ABF and (c) contrast inverted ABF (I-ABF) images. Unit cell averaged (d) ADF, (e) ABF and (f) I-ABF images. The yellow circles in (f) roughly highlighting the position of atom columns. The O columns slightly shift toward right-bottom corner. Intensity profiles of raw experimental and Gaussian fitted (g) ADF and (h, i) ABF images from the regions highlighted in (a) and (b).

To elucidate such a subtle displacement, all the atom columns in both ADF and ABF images are fitted to two-dimensional (2D) Gaussian function using a home-developed MATLAB code [3, 39-43, 45] to minimize the effects of scan noise. The 2D Gaussian peaks rather than the apexes are used to precisely determine the atom positions with sub-pixel precision (~ 0.5 pm precision). Note that we used raw atomic-resolution STEM images without any post-filtering. Intensity profiles in Fig. 1g-i show that the fitted 2D Gaussian profiles reproduce the experimental intensity profiles.

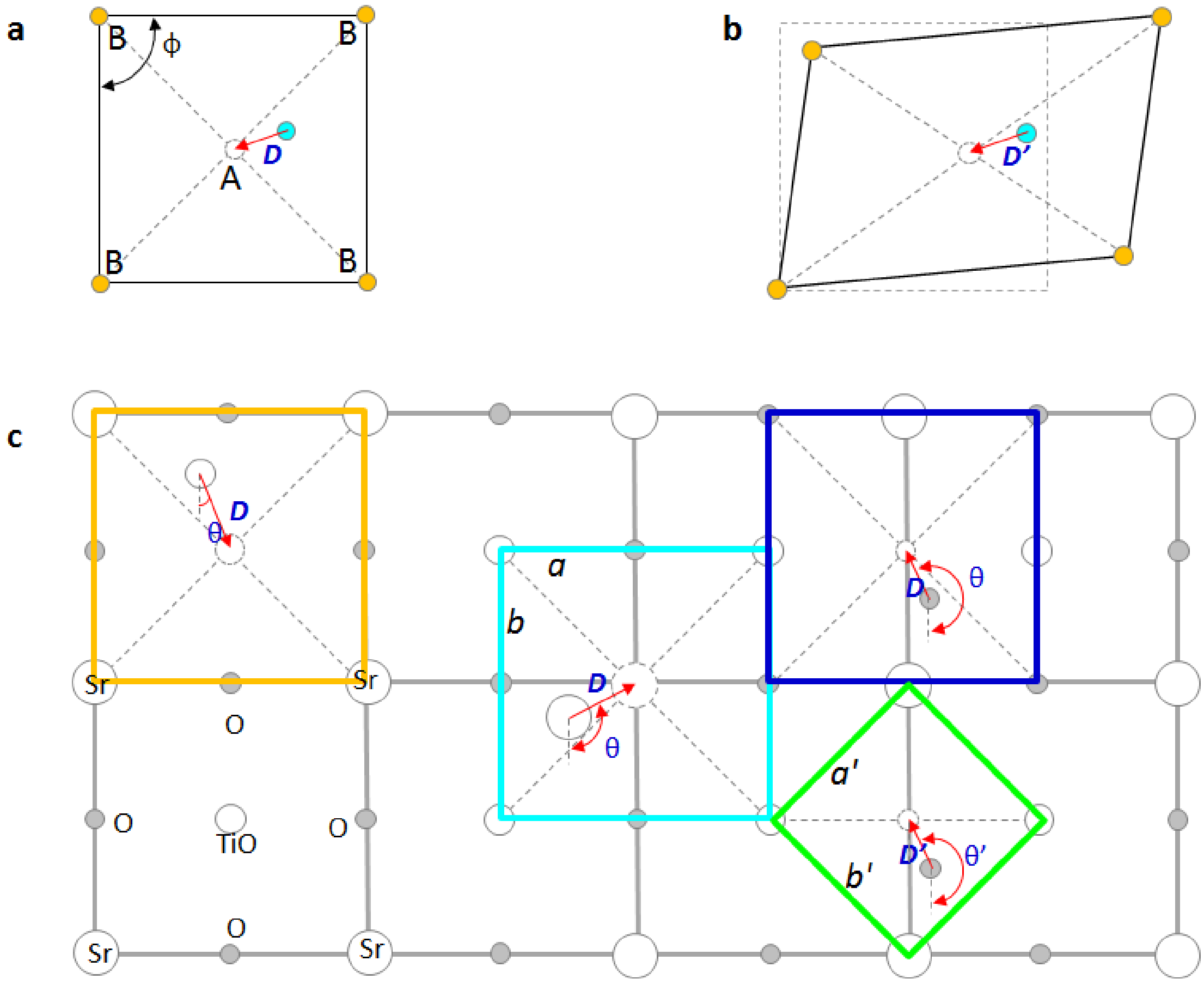

**Figure 2.** Schematic showing relative distance measurement in STEM images to separate the tilt effect from the specimen drift. The STO is viewed along [001] direction. (a) No specimen drift (no image distortion). The displacement vector of A column relative to the center of four nearest B columns is ***D***. (b) With specimen drift (image distortion), B column sub-lattice is not square anymore. The displacement vector of A column relative to the center of four nearest B column sub-lattice is ***D'***. (c) Schematic showing four types of sub-lattice configuration in STEM images. Orange: the displacement of TiO column relative to the center of nearest Sr columns. Cyan: the displacement of Sr column relative to the center of nearest TiO columns. Blue: the displacement of O column relative to the center of neighboring O columns. Green: the displacement of O column relative to the center of neighboring TiO and Sr columns. All the calculated displacements vectors are pointing from the position of atom column to the geometry center of nearest reference columns.

For the evaluation of the specimen tilt, we here use the displacement vector of A-type column relative to the center of B-type sub-lattices. At this geometric configuration as illustrated in Figs. 2a and b, we can reasonably ignore a small constant specimen drift because of subtraction process. In the case of drift-free, the displacement vector is $\boldsymbol{D}=\boldsymbol{P_A}-\boldsymbol{C_B}$, where $\boldsymbol{P_A}$ is the fitted atom position of A-type column and $\boldsymbol{C_B}$ is the center position of B-type sub-lattices. The specimen drift velocity $\boldsymbol{V}$ is assumed to be constant under a short period of recording one-unit cell rows (~160 ms). The pixel corresponding to A-type column position is recorded at the time $t_A$, and the pixels for four B-type column positions are recorded at $t_{B1}$, $t_{B2}$, $t_{B3}$, and $t_{B4}$. Approximately $t_A=(t_{B1}+t_{B2}+t_{B3}+t_{B4})/4$ for the single crystal STO viewing along [100]. In the case of a small drift, the displacement vector becomes $\boldsymbol{D}'=\boldsymbol{P_A}'-\boldsymbol{C_B}'$, where the position of A-type column is $\boldsymbol{P_A}'=\boldsymbol{P_A}+\boldsymbol{V}t_A$, and the center position of four B-type columns is $\boldsymbol{C_B}'=\boldsymbol{C_B}+\boldsymbol{V}(t_{B1}+t_{B2}+t_{B3}+t_{B4})/4$ if the small drift is a constant. In this case, $\boldsymbol{D}'$ is equivalent to $\boldsymbol{D}$, indicating such relative displacement vector is independent of the specimen drift.Therefore, the measured value from experimental STEM images should be mainly due to the specimen tilt. We consider the following displacement vectors with four types of sub-lattices is given in Fig. 2c: Sr *vs* TiO (cation sub-lattice), TiO *vs* Sr (cation sub-lattice), O *vs* O (anion sub-lattice), and O *vs* Sr/TiO (mixture sub-lattice) columns.

In practice, the specimen drift vector is not an exact constant even within a short dwell time, consisting of both '*low*' and '*high*' frequency components. Though the low frequency component can be negligible as a constant for the displacement vector, the high frequency component still exists in STEM images and leads to variable relative displacements in each unit cells. However, the high frequency component is noise-like and we will treat it as a scan noise hereinafter. Although the low frequency component causes a significant change in lattice including lattice aspect ratio and angle between lattice vectors in STEM images (e.g., A-A distance, B-B distance, O-O distance, A-A-A angle etc.), the specimen tilt does not because the atom shift induced by the specimen tilt is exactly the same for the crystallographically same atom columns, and therefore the lattice aspect ratio and lattice vectors angle (A-A-A angle but not A-B-A angle) should be irrelevant with specimen tilt at all. In this regard, the specimen tilt effect and low frequency component of the specimen drift can be reasonably separated, while the effects of high frequency of the specimen drift (scan noise) can be minimized by the 2D Gaussian fitting atom positions and thus ignored for the distance measurements in our study.

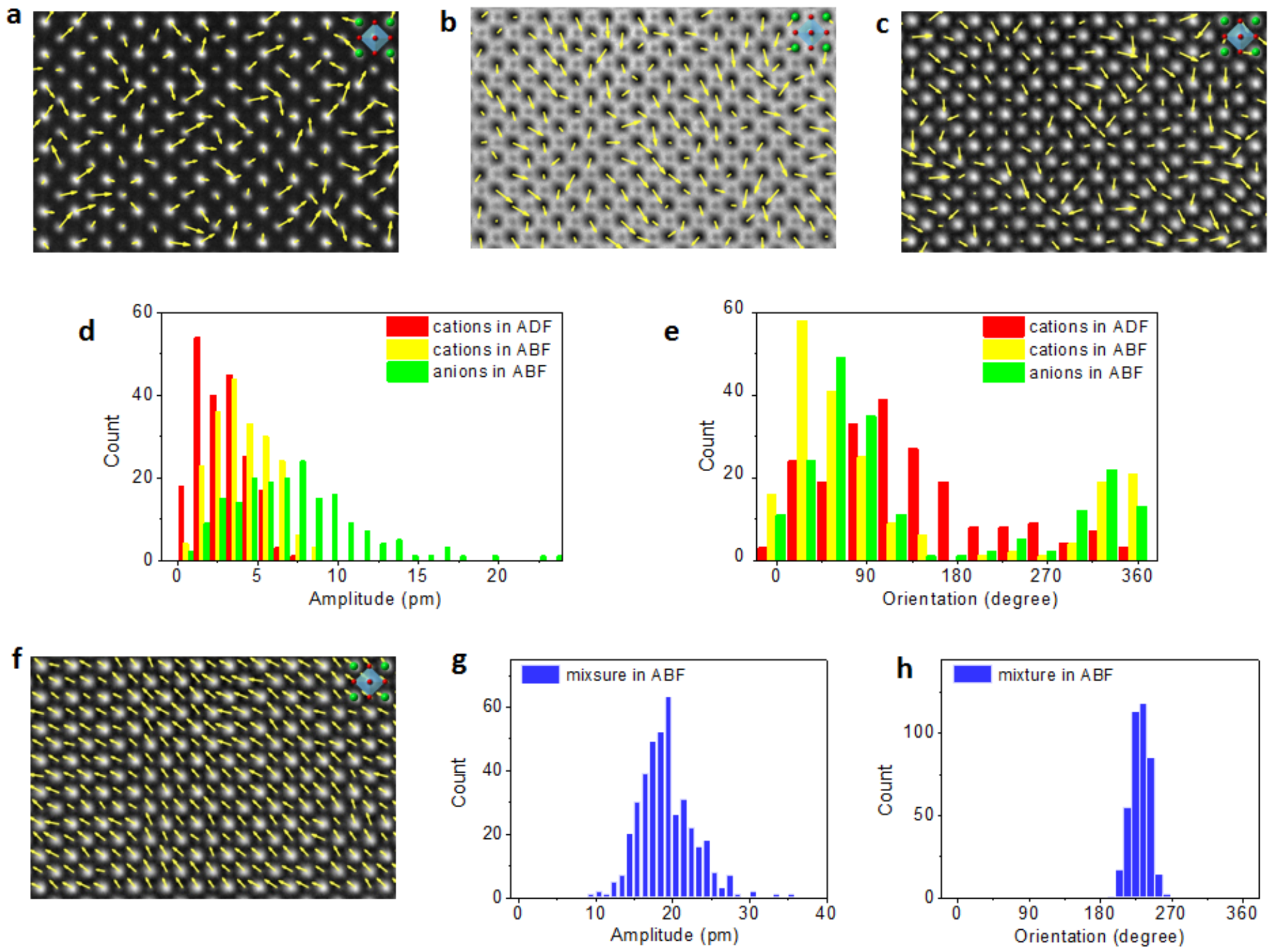

**Figure 3.** Displacement vectors calculated from experimental images. Maps of relative displacement vectors between (a) Sr and TiO columns in ADF, (b) Sr and TiO columns in ABF and (c) O and O columns in I-ABF images. Histogram distribution of (d) amplitude and (e) orientation of displacement vectors. Red: cation sub-lattice in ADF image. Yellow: cation sub-lattice in ABF image. Green: anion sub-lattice in I-ABF image. (f) Map of displacement vectors between O and Sr/TiO columns in I-ABF. Histogram distribution of (g) amplitude and (h) orientation of displacement vectors between O and Sr/TiO.

## 3. Results

### 3.1. Experimental analysis

Figs. 3a and b show the displacement vector maps of cation sub-lattices (TiO column that is a Ti-O-Ti-O-Ti…chain along the electron beam direction is considered as a cation column hereinafter) in ADF and *I*-ABF images, respectively. Fig. 3c is the vector map of anion sub-lattice in *I*-ABF image. The amplitude and orientation of the displacement vectors are summarized in histograms in Figs. 3d and e, respectively. Note that although we have tried our best to tune the specimen zone axis, a tiny crystal tilt that below the detection limit still exists during operation. The average amplitude of the displacement vectors in cation sub-lattices is (2.9±1.5) pm (red) for ADF image and (4.0±1.7) pm (yellow) for ABF image. The measured deviation in ADF image is slightly better than that of ABF image, because ADF image has higher signal-to-noise-ratio (SNR) than that of ABF. For anion sub-lattice in *I*-ABF image, the mean value is (7.0±3.8) pm (green) and the standard error becomes larger than that in the case of cation sub-lattice. This lower precision is attributed to the lower SNR in contrast and fewer pixels for fitting O atom columns. Fig. 3e shows that the displacement vectors in ADF image are slightly random (broader distribution) than those in *I*-ABF image because the former has higher SNR (Note that the orientation of displacement vectors should be random if there is no displacement at all).

In contrast, the specimen tilt induced displacement of anion columns respect to the center of the nearest cation columns is clearly recognizable in Fig. 1(c), which is confirmed by the calculated vector map in Fig. 3f. The amplitude (19.0 ± 3.5) pm in Fig. 3g is much larger than those calculated in Fig. 3d, suggesting the displacement is much higher than that induced by the scan noise and scan distortion. The orientation of vectors is highly ordered pointing to upper-left direction as shown in Fig. 3h, contrasting to approximately random vectors in Fig. 3c. Such artifact leads to significant mis-interpretation of microstructure and properties because the oxygen octahedral distortion/shift in complex oxides accounts for vast functions. It is noteworthy that 19 pm cation displacement relative to the anion in perovskite is the same order of the spontaneous polarization in perovskite ferroelectrics, i.e., 19 pm corresponds to ~19 $\mu C/cm^2$ for displacement between Pb and O in $PbTiO_3$ and ~ 32 $\mu C/cm^2$ for displacement between Ti and O based on the Born effective values from literature [50].

The low frequency component of the sample drift mainly causes the change in lattice constant, lattice ratio and the angle between lattice vectors in STEM images. For each sub-lattice configurations, the bond length *a*, bond angle between *a* and *b*, and lattice ratio *a*/*b* are calculated in Fig. 4. The standard deviation of the measured lattice constant is 3.3 pm (0.83% of lattice constant) for cations sub-lattice, while for anions sub-lattice this value, 6.0 pm, is slightly larger due to the lower SNR as discussed above. The lattice ratio *a*/*b* is 1.008 ± 0.012 (corresponding to 3.2±4.8 pm in distance) and the bond angle is (89.6 ± 0.65)° (corresponding to 2.8±4.4 pm in distance) for cations sub-lattice, whereas the distribution of anion sub-lattice is also slightly broader. The relative large standard deviation of lattice ratio and angle suggests a non-uniform lower frequency distortion during recording the entire image. For the O *vs* Sr/TiO configuration (mixture sub-lattices), the bond length *a'*, bond angle between *a'* and *b'*, and ratio *a'/b'* have no obvious difference from the anion sub-lattice. Overall, the sample drift induced fluctuation in distance is on the level of ~3.2 pm that is relatively small compared to the tilt induced displacement ~19 pm. Such small specimen drift effect is profit from the use of sequential imaging [49].

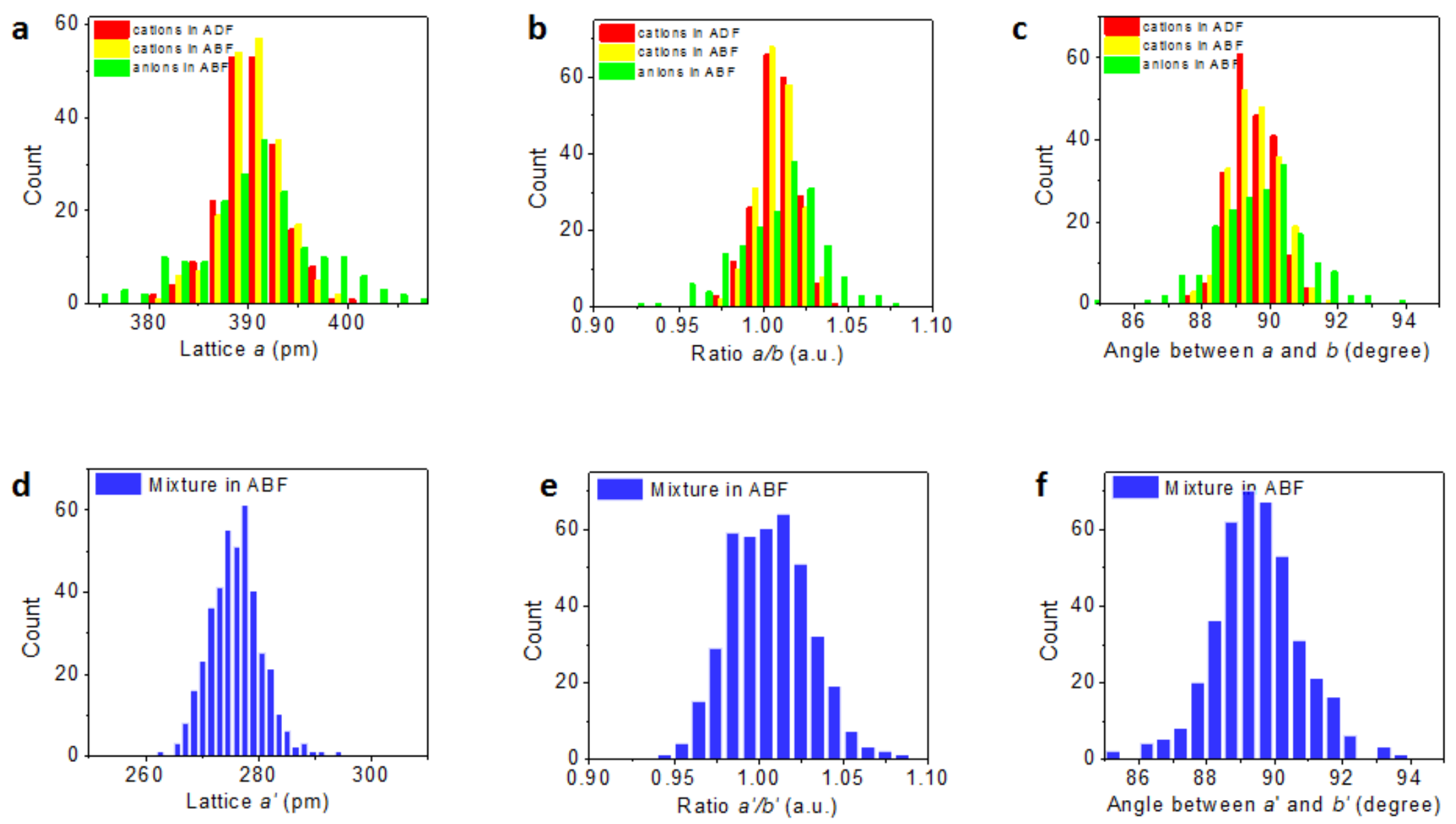

**Figure 4**. Calculation of image distortion. Red: cation sub-lattice in ADF image. Yellow: cation sub-lattice in I-ABF image. Green: anion sub-lattice in I-ABF image. Blue: mixture sub-lattice in I-ABF image. (a) The histogram distribution of lattice *a*. The mean value is normalized to lattice constant 0.3905 nm. (b) The histogram distribution of lattice ratio *a/b*. (c) The histogram distribution of the angle between *a* and *b*. (d) The histogram distribution of lattice *a'* in O *vs* Sr/Ti configuration. The mean value is normalized. (b) The histogram distribution of lattice ratio *a'/b'*. (c) The histogram distribution of the angle between *a'* and *b'*. All the labels *a*, *b*, *a'* and *b'* are listed in Fig. 2c.

### 3.2. Simulation.

It is difficult to distinguish the effects of specimen mistilt from the optical misalignments in STEM. In order to elucidate quantitative information of the net specimen mistilt, a series of STEM images are simulated. Fig. 5a shows the tilt dependence of the simulated aberration-free ADF and *I*-ABF images of $SrTiO_3$ with 25 nm thick, where we slightly mistilt the crystal along the [α ?? 1] direction from the [001] axis and the parameters of α and ?? are the mistilt angles in mrad (the reason we choose 25 nm as thickness is discussed below). No significant atom position shift is observed in the ADF images with α, ??≤10 mrad, while severe displacement appears in the *I*-ABF images with mistilt α, ?? ≥ 6 mrad. Although the respective cation and anion sub-lattices still remain square, their relative positions change dramatically.

Once the mistilts of α, ?? are larger than 10 mrad, the position of anions becomes undistinguishable. To estimate the displacement vectors in these simulated images, we find the atom positions by fitted to 2D Gaussian function in the same manner.

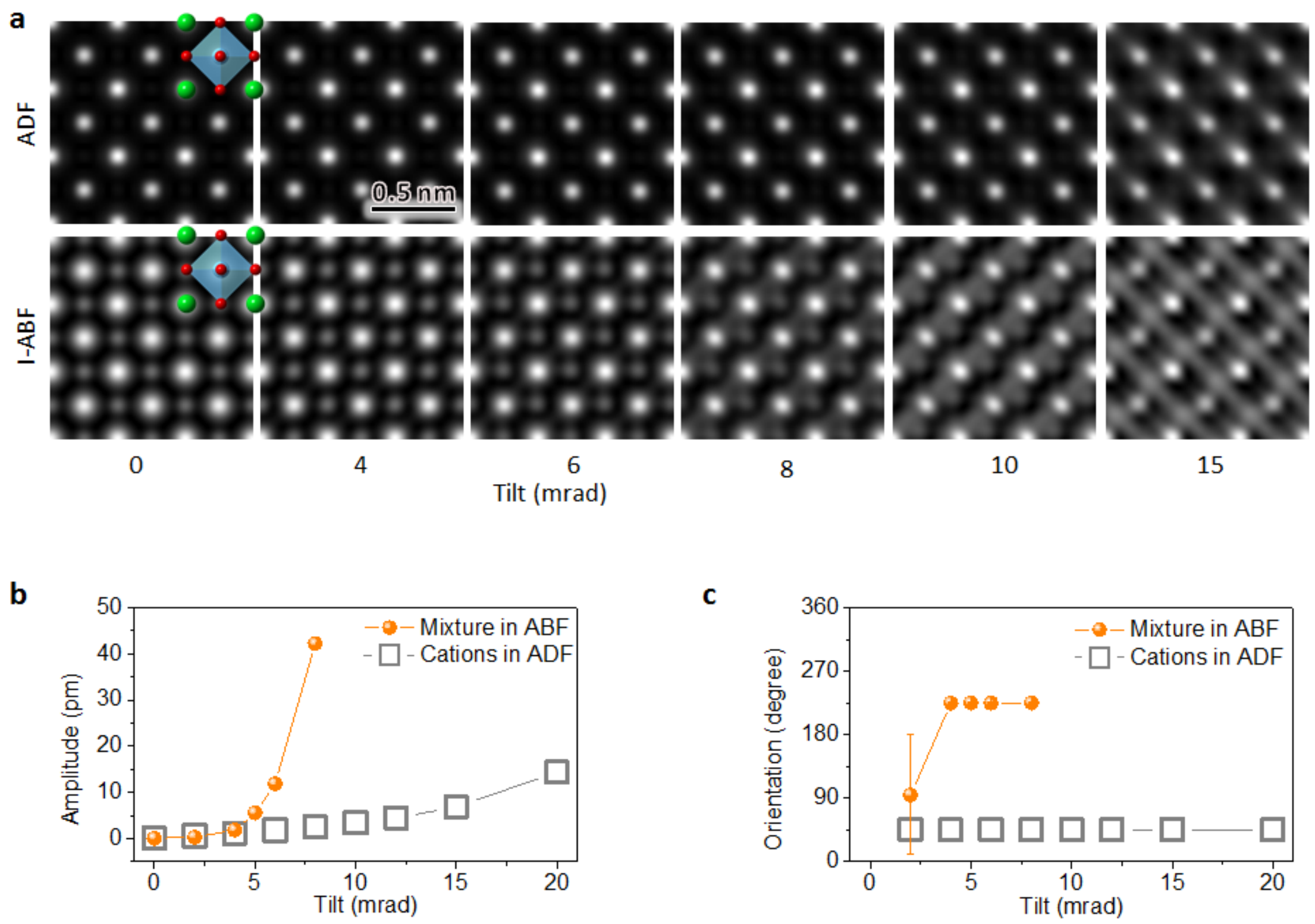

**Figure 5.** Tilt dependence of displacement in STEM images of single crystal STO with thickness of 25 nm and defocus of 0 nm. (a) Simulated ADF and I-ABF images with various tilt: [0 0 1], [0.004 0.004 1], [0.006 0.006 1], [0.008 0.008 1], [0.01 0.01 1] and [0.015 0.015 1]. (b) Amplitude and (c) orientation of displacement vectors calculated from the simulated ADF and I-ABF images.

The amplitude of displacement between Sr and TiO columns in the simulated ADF images shown in Fig. 5b is calculated to be 3.3 pm with α=10 mrad, which is on the same level of the experimental scan noise and distortion. Similar result is obtained from the simulated ABF image, suggesting no obvious dependence of artificial displacement in the cation sub-lattices at the collection angle of detector. However, in ABF images, the displacement between cation and anion changes dramatically with specimen tilt. Only when the tilt α, ??≤4 mrad, the displacement is small enough (≤ 2 pm). Once the misalignment increases to 6 mrad, the displacement reaches 11.9 pm that is much higher than that of scan noise level. With α=??=8 mrad, the displacement is 42 pm which is equivalent to ~11% lattice constant. We note that, in the case without specimen tilt, the "background displacement" (systematic error: fitting noise) is only 0.027 pm for the simulated ADF and 0.048 pm for the simulated ABF, indicating the fitting noise is negligibly small compared to the mistilt effect. In Fig. 5c, the displacement vectors in cation sub-lattice and mixture sub-lattice are orientated in different directions. With mistilt > 2 mrad, the vectors calculated from mixture sub-lattice are orientated in the same direction, indicating the presence of significant artificial displacements.

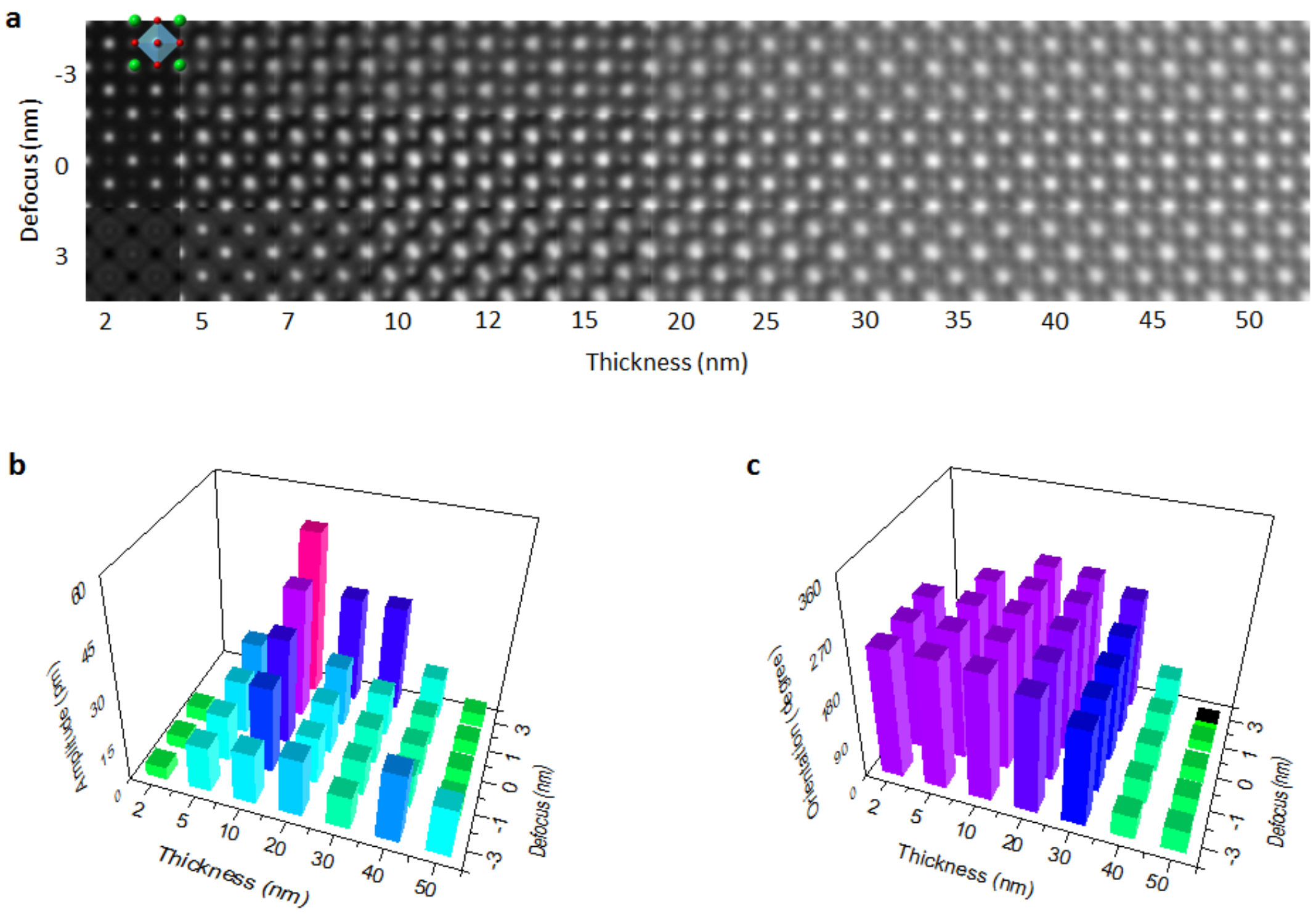

**Figure 6.** Thickness and defocus dependence of displacement vectors in single crystal STO with zone axis [0.008 0.005 1]. (a) Simulated I-ABF images with different thickness and defocus. (b) Amplitude and (c) orientation of displacement vectors between O and Sr/TiO columns.

The displacements between different atom species also rely on the defocus and thickness of specimen, because the probe propagation with the tilted crystal shows a dechanneling to the other atomic columns [28]. Fig. 6a is the simulated *I*-ABF images of $SrTiO_3$ with slightly off-axis [α ?? 1] with (α, ??) = (8, 5) mrad. It clearly shows the displacements strongly depend on mistilt angles. Typical frames with various thickness and defocus are selected for 2D Gaussian fitting and the displacement vectors are estimated. The calculated amplitude and orientation of displacement are summarized in Fig. 6b and c, respectively. By changing either thickness or defocus, the dependence behavior of amplitude is rather complicated. In our simulations, the relative displacements (>50 pm) for the mixture sub-lattice are maximized in the thickness range of 10-15 nm. The amplitude of displacements in ABF simulations also changes as a function of defocus. The dependence of orientation on thickness is not monotonous. When the thickness increases from 20 to 40 nm, the direction of displacement vectors becomes opposite. The defocus does not influence the orientation significantly. Furthermore, for ADF images, increasing convergence angle has been proposed previously to reduce the impact of the tilt effect [21]. To investigate validity of larger illumination angles to the ABF imaging, we performed systematic image simulation with several convergence angles as well as collection angles (in order to match the 'hollow-cone illumination' conditions [11] to optimize the ABF contrast [8]). The displacements between cation and oxygen are plotted in Fig. 7. For the convergence angle of 10 mrad, the displacement in the mixture sub-lattice becomes over 30 pm only with 0.5 mrad mistilt. For the larger convergence angles, the displacement can be sufficiently reduced with a small mistilt. Therefore, it should be helpful to use a small illumination angle to detect the mistilt during specimen alignment and a large illumination angle for ABF imaging to sufficiently reduce mistilt effects [49, 51].

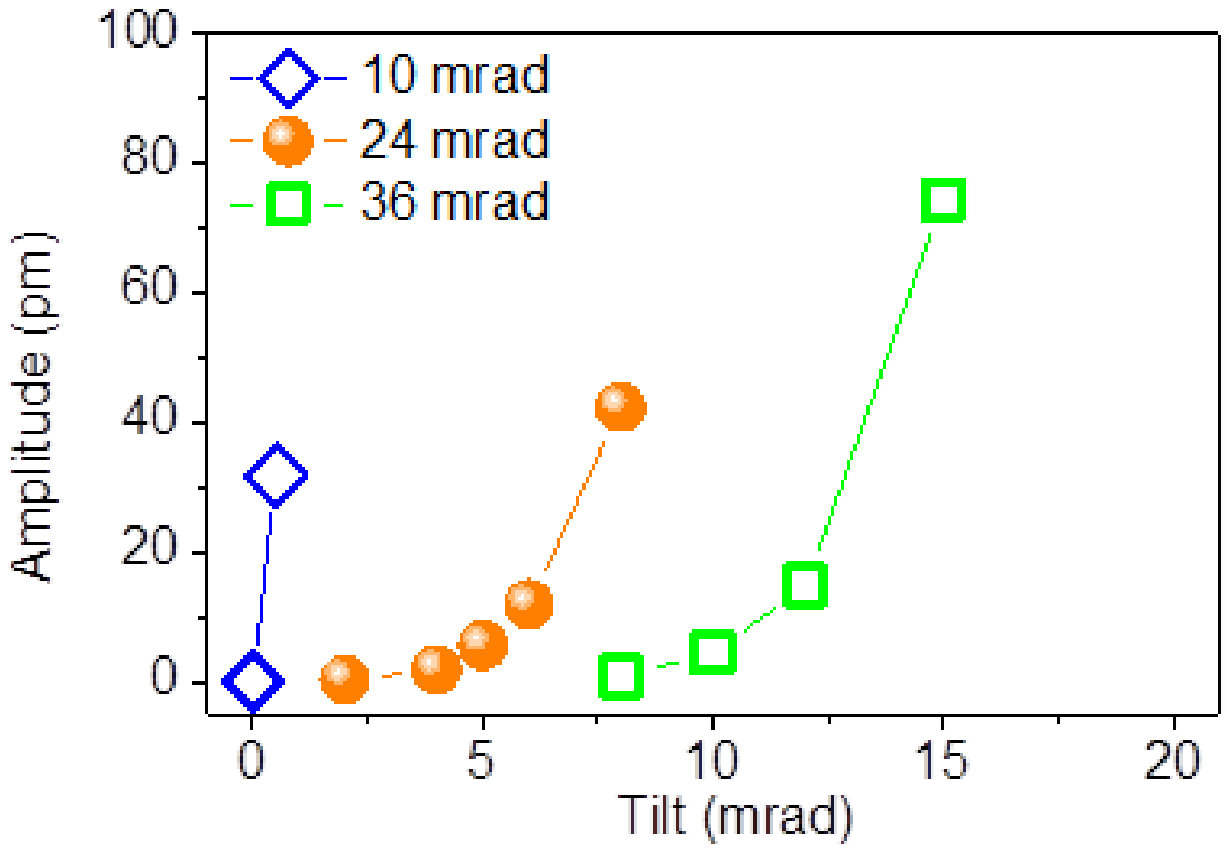

**Figure 7.** Amplitude of displacement vectors between cation and oxygen columns in the simulated ABF images with thickness of 25 nm, defocus of 0 nm and various convergence and collection angles. Blue diamond: 10 mrad and [5, 10] mrad for convergence semi-angle and collection angles. Orange ball: 24 mrad and [12, 24] mrad. Green square: 36 mrad and [18, 36].

Since the measured displacement vectors in STEM images are affected by crystal tilt, specimen thickness, defocus, and convergence angle, we could find the practical parameters that can reproduce the artificial displacement in experiments with the aid of image simulations. Fig. 8a and b are simulated *I*-ABF image and corresponding vectors map along the zone axis of [α ?? 1] with (α, ??) = (8, 5) mrad, thickness of 25 nm, defocus of 0.5 nm and the convergence semi-angle of 24 mrad.

On the basis of the systematic investigations in simulated images, we found that the experimental data in Fig. 3g and h are in good agreement with the histograms of the amplitude and the orientation of displacement vectors in Fig. 8c and d. Moreover, the simulated ABF intensity profiles with 25 nm thickness are matched with the experimental intensity profiles (Figs. 8e, f) very well.

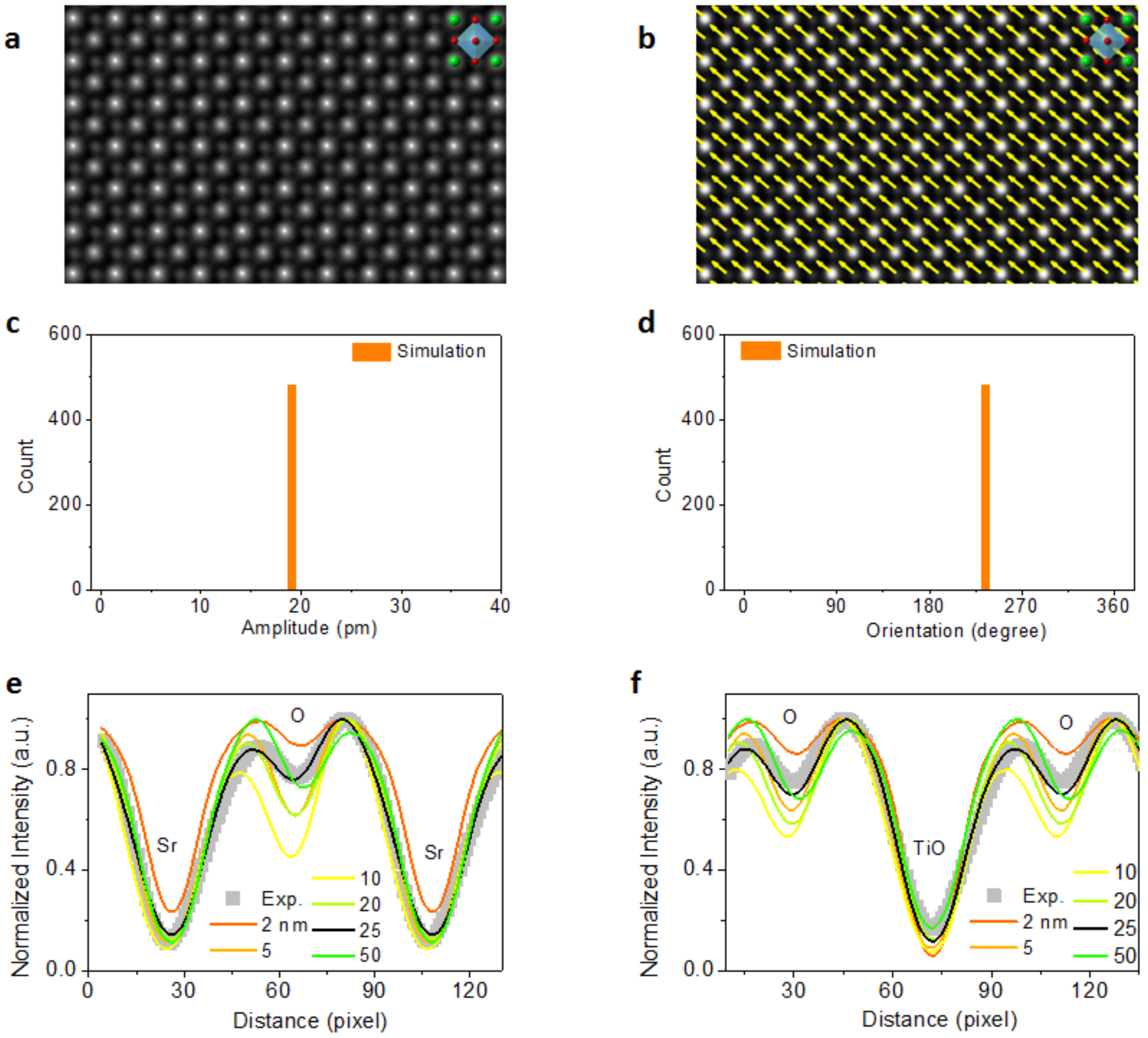

**Figure 8.** Simulation of STO with viewing direction of [0.008 0.005 1], thickness of 25 nm and defocus of 0.5 nm. (a) Simulated I-ABF image. (b) Corresponding map of displacement vectors between O and Sr/TiO columns in the simulated I-ABF image. Histogram distribution of (c) amplitude and (d) orientation of displacement vectors. (e, f) Thickness-dependent intensity profiles showing 25 nm (black curves) is close to the experimental data (grey squares).

However, it should be noted that despite good agreement between experimental data and simulation, the experimental conditions are likely (more or less) different from simulation parameters due to the presence of coma, astigmatism, misalignments of aperture and detector, and therefore the simulated results can be considered as equivalent ‘effective tilt’ effects, which will discuss later. Nevertheless, the simulation can help us to effectively determine the specimen tilt and thus correct the artificial displacements.

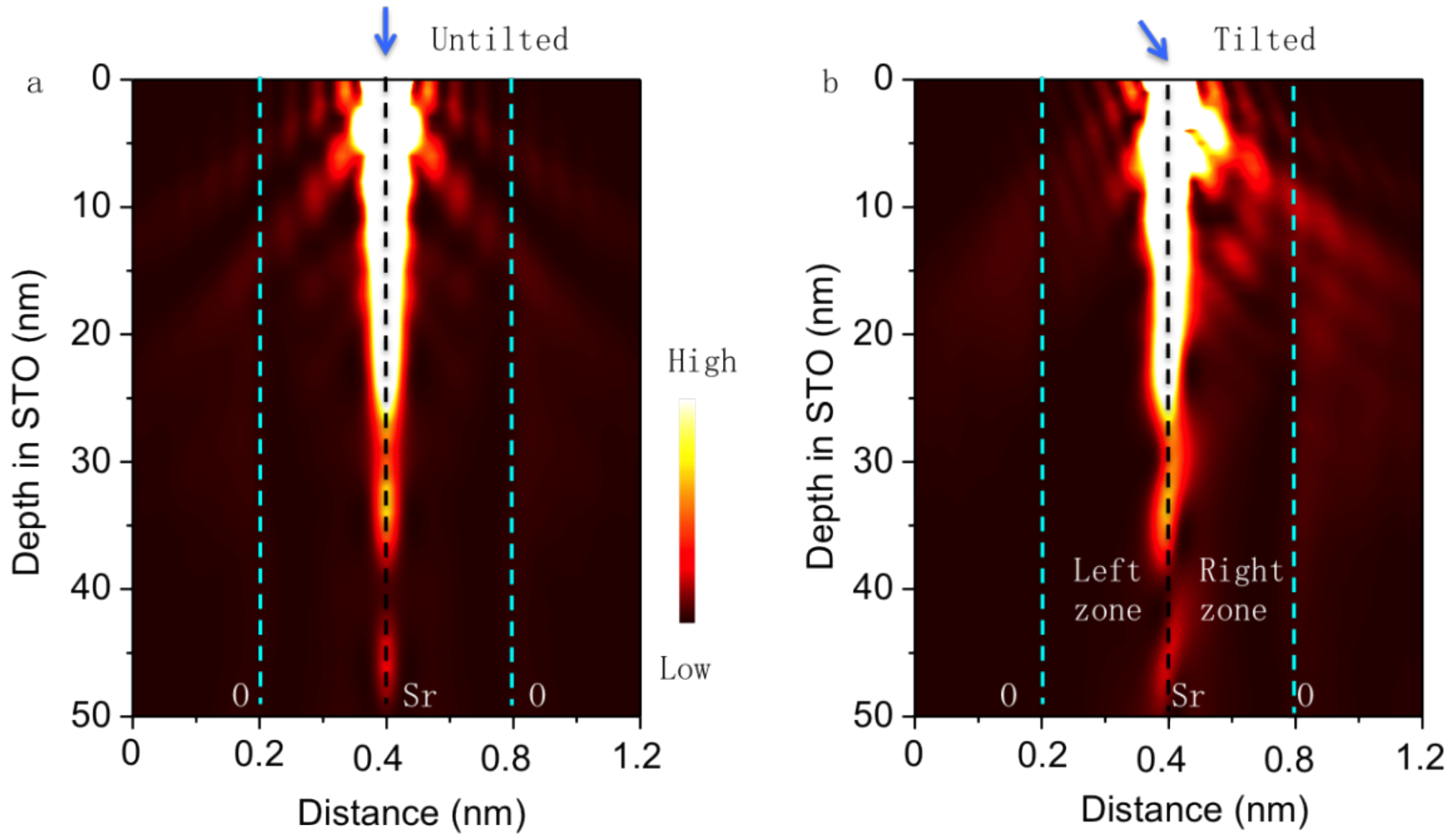

**Figure 9**. Calculated beam propagation in the STO with convergence angle of 25 mrad. (a) Zone axis is [0 0 1]. (b) Zone axis is [0.008 0 1]. Vertical black dashed lines highlight the Sr position. Vertical cyan dashed lines highlight the O positions. In the tilted case, complex contrast exist in the right side of Sr and the left side is relatively clean due to the asymmetric dechanneling behavior.

## 4. Discussion

To reveal why the atom positions in ABF image is sensitive to the specimen tilt, we calculated the probe intensity propagation in the $SrTiO_3$ in Fig. 9, where the electron beam is focused on the entrance surface of the Sr column. In the case of tilted specimen in Fig. 9b (α=8 mrad), the electron probe propagation along the Sr column periodically oscillates through the thickness. Between the Sr and O columns, complex contrast exists due to the dechanneling of the incident beam from Sr to the neighboring O columns. The resultant contrast by the dechanneling becomes no longer symmetric along the atom column direction (i.e., lower contrast in the left gap between Sr and O column and higher contrast in the right gap). The atom positions are controlled by both the channeling electrons at the atom column and the de-channeling electrons from the neighboring columns. For the contrast at lighter O columns, the amount of de-channeling electrons from the neighboring Sr and TiO columns contribute significantly to the entire signal and thus the measured O position in the ABF is strongly affected by the dechanneling electrons. In the case of asymmetric dechanneling behavior of Sr column along the tilt direction in Fig. 9, the position of right O column appears to shift close to Sr column whereas the left one shifts away, leading to a displacement between Sr and the center of neighboring O columns. In contrast, once we move the electron probe from Sr to O column, because compared to the channeling electrons of Sr column the amount of de-channeling electrons from the neighboring O columns is small [8], the position of Sr column is still dominated by the channeling electrons of Sr and thus the position of Sr is not significantly changed in the ABF image. Overall, the position of heavy atom columns is less sensitive to the specimen tilt [21], as shown in Fig. 5a.

In the case of specimen tilt, during image recording we usually deliberately introduce coma and astigmatism to partly compensate the tilt effect thus to get better visual appearance of the atom shape. Therefore, the specimen tilt can not only directly affect the positions of atom columns in STEM images but also indirectly influence the contrast via introduction of aberration. To evaluate the effects of aberration, the simulated ABF images with different $A_1$, $A_2$, $B_2$ and $C_3$ are shown in Fig. 10. When the $A_2$ or $B_2$ is smaller than 50 nm, the displacement is less than 5 pm that is at the level of experimental error such as scan noise or scan distortion. Note that, in our experiments, we use the automated software to set up the values of $A_2$ and $B_2$ less than 20 nm at amorphous regions [52]. In this regard, the net effects of aberration are negligible on the quantitative position analysis. Consequently, most of the quantitative discussion below is based on the direct tilt effect in the simulation, but these conclusions should be qualitatively applicable to experimental data.

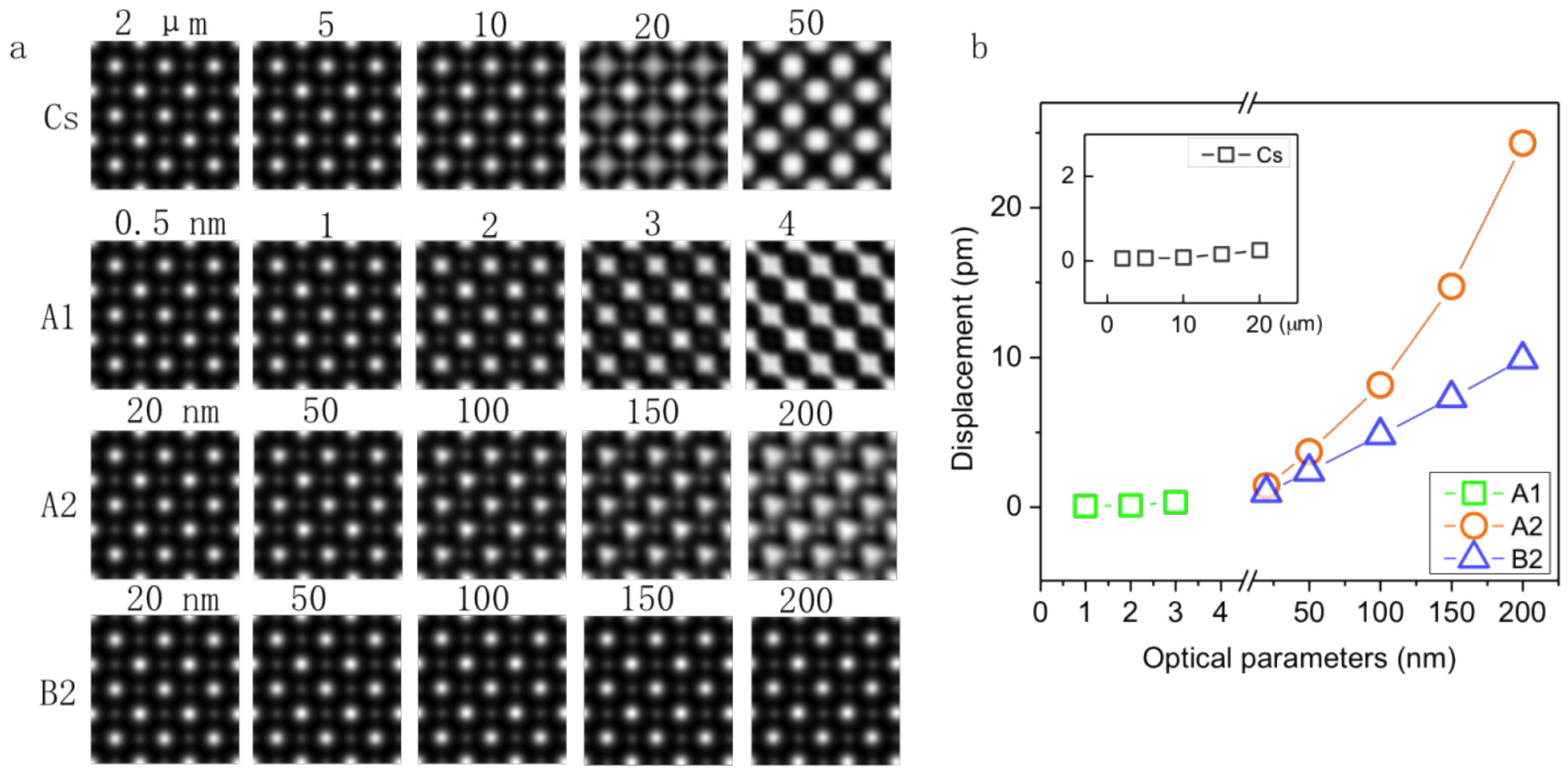

**Figure 10**. Aberration dependence of displacement in single crystal STO with zone axis [0 0 1]. (a) Simulated I-ABF images. (b) Amplitude of displacements in the I-ABF images with aberration.

To quantitatively discuss the crystal tilt effect on precise position measurement, we define an 'effective sub-lattice', within which calculations between any atom columns remain effective. In the case of perfect zone axis ($\alpha$=0), the entire lattice is an effective sub-lattice. With small tilt $\alpha \leq 4$ mrad for ~ 25 nm thick STO, since all types of displacements are small enough (< 2 pm), the entire lattice can still be treated as an effective sub-lattice. Once the mistilt is $\alpha > 5$ mrad, the displacement between O and cations columns is ~11.9 pm for $\alpha$=6 mrad that is well above the noise level and scan distortion. Therefore, the anion and cation columns in $SrTiO_3$ are no longer in the same effective sub-lattice, but the Sr and TiO columns can be still considered within the same effective sub-lattice if the tilt is ≤10 mrad. For large crystal mistilt >15 mrad, the Sr and TiO columns are not within the same effective sub-lattice any more. However, respective Sr, TiO and O columns are intrinsically in the same effective sub-lattice regardless of tilt as they have the same shift behavior on tilt, indicating unlike scan noise and distortion the specimen tilt will not affect the measurement of lattice constant, lattice ratio and angle between lattice vectors.

## 5. Conclusions

The effects of specimen tilt on the picometer-scale measurement of atom positions for ABF-STEM images are studied by combing the experiments and simulations. By using the relative distance measurement method, we are able to effectively distinguish the effects of crystal tilt from scan noise and distortion. Thus, it is possible to quantitatively evaluate the specimen tilt effects on the atom position analysis. The main conclusions are given below:

(1) For ABF imaging, small tilt on the order of ~ 6 mrad can cause an artificial displacement 11.9 pm between anion and cation positions under typical experimental conditions for cubic $SrTiO_3$. This value is 3 or 4 times larger than the scan noise and sample drift induced image distortion (3.2 pm), suggesting the specimen tilt is critical for picometre-scale measurement of ABF images because such a small tilt is inevitable during STEM operation.

(2) The tilt-induced artifact relies on the tilt angle, thickness of specimen, defocus and convergence angle. During specimen alignment, changing defocus and using small aperture is helpful to detect and correct the specimen tilt. However, larger aperture is preferable for imaging to minimize the artifact.

(3) Although the residual aberrations can also influence on the atom position analysis, the effects of aberration is difficult to separate from the net mistilt in the experimental images. Generally, small $A_2$ and $B_2$ (< 50 nm) would not cause significant artifact on displacements measurement.

(4) The effective sub-lattices can help us to evaluate the effects of tilt. The effects of small mistilt in ADF images is negligible because the cation columns such as Sr and TiO have similar dechannelling behavior (while the light atoms are invisible). The pure tilt effect has no influence on the lattice constant measurement because the crystallographically same atom columns have the same atom shift.

(5) The tilt induced asymmetric dechannelling along the beam propagation accounts for the atom position shift and artificial displacements.

(6) The relative displacement measurement method that can effectively separate the tilt effect from the scan noise and scan distortion can be used to quantitatively evaluate the tilt effect. Multislice simulation of STEM images can be used to quantitatively estimate tilt effects in the experiments, and thus help us to extract the true atomic configurations.

These findings provide useful insights into quantitative calculations based on STEM images.

**Acknowledgements**

The authors acknowledge Dr. Takehito Seki, Dr. Yeong-Gi So and Dr. Christopher Nelson for their helpful discussion and data processing techniques. This work was supported in part by the Grant-in-Aid for Specially Promoted Research (Grant No. 17H06094) from Japan Society for the Promotion of Science (JSPS), and "Nanotechnology Platform" (Project No. 12024046) from the Ministry of Education, Culture, Sports, Science and Technology in Japan (MEXT). P.G. was supported as a Japan Society for the Promotion of Science (JSPS) fellow for part of this work.